\documentclass[runningheads]{llncs}

\usepackage{amsmath}
\usepackage{multirow}
\usepackage{color}
\usepackage{bm}
\usepackage{subfig}
\usepackage{url}
\usepackage{booktabs}
\usepackage{pgfplots}
\usepackage{xspace}
\pgfplotsset{compat=1.8}

\usepgfplotslibrary{statistics}

\usepackage{xcolor}
\definecolor{darkgreen}{rgb}{0,0.5,0}

\newcommand\extended[1]{#1}

\newcommand{\leanparagraph}[1]{\vspace*{1mm}\noindent\textbf{#1.}~~}

\newcommand{\tkmatch}{\textsc{T2K Match}\xspace}
\newcommand{\tableminer}{\textsc{Table\hspace{0pt}Miner+}\xspace}

\newcommand{\squishlist}{
 \begin{list}{$\bullet$}
  { \setlength{\itemsep}{0pt}
     \setlength{\parsep}{3pt}
     \setlength{\topsep}{3pt}
     \setlength{\partopsep}{0pt}
     \setlength{\leftmargin}{1em}
     \setlength{\labelwidth}{1em}
     \setlength{\labelsep}{0.5em} } }
\newcommand{\squishend}{
\end{list}}
\newcommand{\cL}[0]{\ensuremath{\mathcal{L}}}

\newcommand{\cF}[0]{\ensuremath{\mathcal{F}}}
\newcommand{\cE}[0]{\ensuremath{\mathcal{E}}}

\newcommand{\cK}[0]{\ensuremath{\mathcal{K}}}

\newcommand{\row}[0]{\ensuremath{\rho}}
\newcommand{\col}[0]{\ensuremath{c}}
\newcommand{\cR}[0]{\ensuremath{\mathcal{R}}}
\newcommand{\rel}[0]{r}
\newcommand{\sub}[0]{s}
\newcommand{\obj}[0]{o}

\date{}

\begin{document}

\title{Extracting Novel Facts from Tables for Knowledge Graph Completion\extended\\(Extended version)}

\author{
Benno Kruit\inst{1}\inst{2} \and Peter Boncz\inst{1} \and Jacopo Urbani\inst{2}\\
}
\institute{
Centrum Wiskunde \& Informatica, Amsterdam, The Netherlands
\and
Dept. of Computer Science, Vrije Universiteit Amsterdam, The Netherlands\\
\email{\{kruit, p.boncz\}@cwi.nl; jacopo@cs.vu.nl}
}

\maketitle

\begin{abstract}

We propose a new end-to-end method for extending a Knowledge Graph (KG) from
tables. Existing techniques tend to interpret tables by focusing on information
that is already in the KG, and therefore tend to extract many redundant facts.
Our method aims to find more novel facts. We introduce a new technique for table
interpretation based on a scalable graphical model using entity similarities.
Our method further disambiguates cell values using KG embeddings as additional
ranking method. Other distinctive features are the lack of
assumptions about the underlying KG and the enabling of a fine-grained tuning of
the precision/recall trade-off of extracted facts. Our experiments show that our
approach has a higher recall during the interpretation process than the
state-of-the-art, and is more resistant against the bias observed in extracting
mostly redundant facts since it produces more novel extractions.  \end{abstract}


\section{Introduction} 

\leanparagraph{Motivation} Much of the world's information exists as tabular
data. These are available as HTML tables on web pages, as spreadsheets, or as
publicly available datasets in many different formats. There has been more than
a decade of research in recognizing, cleaning and capturing these so-called {\em
web tables}~\cite{cafarella2018}.  Because of their  relational nature, such
large collections of web tables are suitable for  supporting table
search~\cite{yakout2012infogather} or for answering specific  factual
queries~\cite{Sun2016a}.  In certain web tables, the rows describe attributes or
relationships of entities. This makes them suitable sources for extending the
coverage of Knowledge Graphs (KGs), which is a task known as {\em KG completion}.

\leanparagraph{Problem}  In order to perform KG completion from web tables, we
must first align their structure and content with the KG, a problem broadly
referred to as \emph{table interpretation}. Table interpretation has been the
subject of several prior works \cite{Limaye2010AnnotatingAS,Ritze2016Profiling,zhang2017effective,Mulwad2013,Munoz2014,Sekhavat2014KnowledgeBA,Ermilov2016,Ibrahim2016}.
Similar to our research,
these works primarily focus on the interpretation of entity tables, i.e.,
tables where each row describes one entity and columns represent attributes.
In this case, the interpretation process consists of two operations. First,
each row is linked with an entity in the KG, and optionally the entire table
is linked to a class. Then, each column is associated to a KG relation.

After the table is correctly interpreted, we can extract novel triples from the
table and add them to the KG. This last operation is also known as
\emph{slot-filling}, as the empty `slots' in the KG are filled with new
facts~\cite{Ritze2016Profiling}. Table interpretation strongly affects the
quality of slot-filling, since errors in the former can no longer be corrected.
Because of this, state-of-the-art table interpretation techniques (an overview
is given in Section ~\ref{sec:related}) aim for high precision by pruning out
many potential assignments already at early stages. While high precision is
desirable in some contexts (e.g., table search), it has been
observed~\cite{kruit2018extracting} that this strategy leads to a high number of
redundant extractions during slot-filling, since only the assignments to
entities that are well-covered in the KG are retained.

\leanparagraph{Contribution} With the goal of maximizing the number of novel
extractions without sacrificing precision, we present a new method for KG
completion from web tables. In contrast to existing approaches, our method
does not prune out row-entity assignments, but performs the interpretation by
performing inference over all possible assignments using a Probabilistic
Graphical Model (PGM). The PGM uses label similarities as priors, and then
updates its likelihood scoring to maximise the {\em coherence} of entity
assignments across the rows using Loopy Belief Propagation (LBP). Coherence is
not computed using a predefined metric (such as class membership) but is
automatically selected as a
combination of properties that are shared by the entities in the table. This
is a novel feature of our method which makes it capable of working with KGs
with different topologies and/or relations.
Since we use both label similarities and coherence based on
salient common attributes, our method is able to maintain a high accuracy
for the row-entity assignments. At the same time, it is also able to return many
more novel extractions since we did not prune out any assignments.

We also propose an approach to perform slot-filling by disambiguating attribute
cells in a novel link-prediction framework. Our approach makes use of embeddings
of KG entities and relations to improve the quality of the disambiguation
whenever label matching is not sufficient. This furthers our aim to find novel
facts for KG completion.

We compared our method to several state-of-the-art systems. Additionally, we
evaluated the performance of these systems with regard to the redundancy of the
facts that they extract from the tables. Our experiments on popular benchmark
datasets show that our approach yields slightly lower precision, but
significantly higher recall on entity predictions. This leads to many more novel
extractions than what is possible with existing methods. Finally, to test the
scalability of our method we perform a large-scale evaluation on 786K tables
from Wikipedia. An extended version of this paper is available
at~\cite{extended}.



 \label{intro}


\section{Background}
\label{sec:background}

\leanparagraph{KGs} A KG $\cK$ is a repository of factual knowledge that can be
seen as a directed labeled graph where the nodes are entities and the edges
represent semantic relations. We define $\cK$ as a tuple $(\cE,\cR,\cF)$ where
$\cE$ is the set of entities (nodes), $\cR$ is the set of relations, and $\cF$
is the set of facts (edges) in the graph. Each entity is associated to a finite
set of labels $\textsf{Labels}(e)$. We use the notation $\langle
\sub,\rel,\obj\rangle$ to denote a fact in $\cF$ where
$\sub,\obj\in\cE$ and $\rel\in\cR$. Realistic KGs contain facts of
various types: For instance, they either indicate type memberships (e.g.,
$\langle$\texttt{Nether\-lands}, \texttt{type}, \texttt{Country}$\rangle$), or
encode more generic binary relations (e.g., $\langle$\texttt{Amster\-dam},
\texttt{capitalOf}, \texttt{Nether\-lands}$\rangle$), and are normally encoded in
RDF \cite{rdfs}.

\leanparagraph{Table Interpretation} Tables represent an important source of
knowledge that is not yet in the KG. A class of tables that is particularly
useful for enriching KGs is the one that contains \emph{entity tables}, i.e.,
tables where one column contains the name of entities (the \emph{key-column}) and all others contain the
entity attributes. While these tables ostensibly contain structured data, the textual
content of cells and identifiers is created more with the aim of human
interpretation than automatic processing. To capture the semantics of these
tables in a coherent and structured way, it is useful to link their content to
concepts in the KG. We refer to this task as \emph{table interpretation},
mapping each row and attribute column to entities and relations
respectively. These mappings can be computed by determining (1) which entities
in the KG are mentioned in the table, (2) which are the types of those entities,
and (3) which relations are expressed between columns (if any)
\cite{Limaye2010AnnotatingAS,Venetis2011RecoveringSO,Mulwad2013,Ritze2015MatchingHT,zhang2017effective}.
After the interpretation is finished, we can use the mappings to construct facts
for the KG. We call this operation \emph{slot-filling}.

\extended{
\begin{example} Consider an example KG with five entities
    $\{$\texttt{Netherlands}, \texttt{Country}, \texttt{Amsterdam},
    \texttt{City}, \texttt{capitalOf}$\}$ and a table $X$ which contains a row
    $r$ with cell values $r[1]=$ ``Holland" and $r[2]=$ ``A'dam". The first cell
    value should be mapped to the entity \texttt{Netherlands}, while the second
    should be mapped to the entity \texttt{Amsterdam}. The mapping is not
    trivial because a string can map to multiple entities, e.g., ``Holland" can
    refer either to the county or to 19 different cities in the U.S. The task of
    table interpretation consists of disambiguating the correct meaning intended
    in the table. Furthermore, if the other rows in the table also contain
    countries and their capital cities, then the system should infer that all
    these entities are instances of classes such as \texttt{Country} and
    \texttt{City}, and the relation between the columns should be
\texttt{capitalOf}. With slot-filling, our goal is to extract statements like
$\langle \texttt{Amsterdam},\texttt{capitalOf},\texttt{Netherlands}\rangle$ from
the table so that they can be added to the KG.\end{example}
}



\leanparagraph{PGMs} In this paper, we employ Probabilistic Graphical Models
(PGMs) to perform the interpretation. PGMs are a well-known formalism for
computing joint predictions \cite{Pearl1989ProbabilisticRI}. For a given set of
random variables, conditional dependences between pairs of variables are
expressed as edges in a graph. In these graphs, variables are connected if the
value of one influences the value of another. The connection is directed if
the influence is one-way, and undirected if both variables influence each
other.  The behaviour of the influence on every edge is expressed by a
function known as the \emph{potential function}. When performing inference in
a PGM, information from the nodes is propagated through the network using the
potential functions in order to determine the final distribution of the random
variables.

\leanparagraph{KG Embeddings} We also make use of
latent representations of the KG~\cite{nickel2016review}
to filter out incorrect extractions. In particular,
we consider \emph{TransE}~\cite{transe}, one of the most popular methods in this
category.  The main idea of TransE is to ``embed'' each entity and relation into
a real-valued $d$-dimensional vector (where $d>0$ is a given hyperparameter).
The set of all vectors constitutes a model $\Theta$ of $|\cE|d+|\cR|d$ parameters
which is trained so that the distance between the vectors of entities which are
connected in $\cK$ is smaller than the distance between the ones of entities
which are not connected. 

\extended{
Training $\Theta$ is done by minimizing the loss function
\vspace{-2mm}
\begin{equation}
    \cL_\Theta = \sum_{\langle \sub,\rel,\obj \rangle \in \cF} \sum_{\langle
    \sub',\rel,\obj'\rangle \in S_{\langle \sub,\rel,\obj \rangle}} [\gamma + d(
    \mathbf{\sub} +
    \mathbf{\rel},\mathbf{\obj}) -
    d(\mathbf{\sub'} + \mathbf{\rel},\mathbf{\obj'})]_+
\end{equation}
\noindent where: $\mathbf{\sub,\sub',\obj,\obj',\rel}$ are the vectors
associated to the
entities $\sub,\sub',\obj,\obj'$ and type $\rel$ respectively; $\gamma\geq 0$ is
an
hyperparameter that defines the minimum acceptable margin; $d(\cdot)$ is a distance
function (typically the $L_1$ norm), $[x]_+$ returns the positive part of $x$,
and $S_{\langle \sub,\rel,\obj \rangle}=\{\langle \sub,\rel,\obj' \rangle \mid \langle
\sub,\rel,\obj'\rangle \notin \cF\} \cup \{\langle \sub',\rel,\obj \rangle \mid \langle \sub',\rel,\obj
\rangle \notin \cF\}$, i.e., it is a set of ``corrupted'' facts which are not in
$\cK$. Once training is completed, the model can be used to perform link
prediction, i.e., to estimate the likelihood of unseen facts: if the distance of
these facts is small, then these are more likely to be true.
}
 \label{background}


\begin{figure*}[t]
\centering
\makebox[\textwidth]{\includegraphics[width=\textwidth]{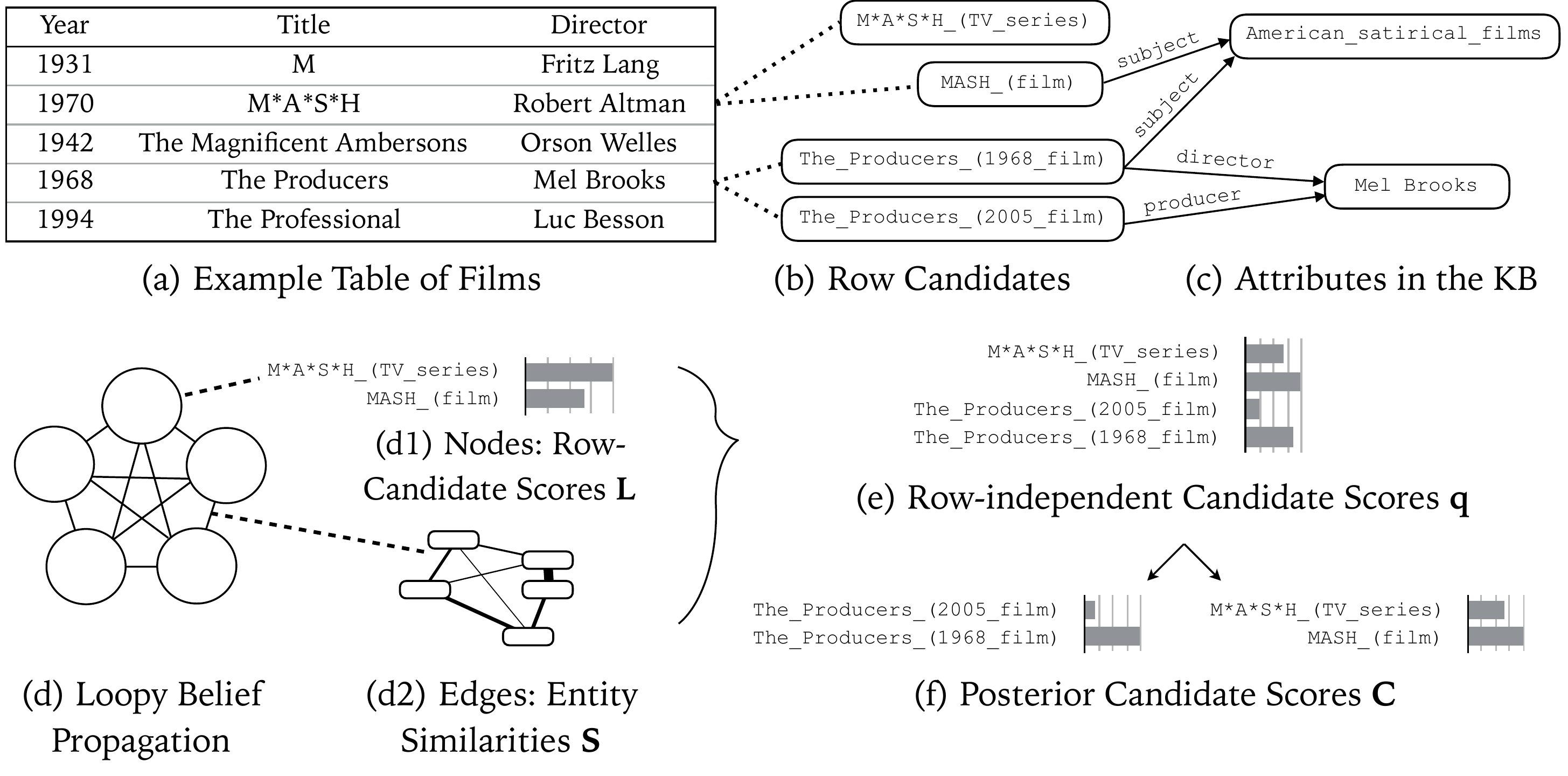}}
\caption{Schematic representation of our method.
}
\label{fig:method}
\end{figure*}

\section{Table Interpretation} 
We introduce our method for performing table
interpretation. Fig.~\ref{fig:method}
shows the computation that takes place during the interpretation,
using table (a) as a motivating example. In this case, the key-column
is the second one (``title'') but its content is ambiguous since the values can
refer to movies, TV series, or books. For instance, the second row can refer to
the TV serial \texttt{M*A*S*H} or to the movie \texttt{MASH}, as is shown in
Fig.~\ref{fig:method}b.
The goal of this task is to map as many rows $\row$ as possible to corresponding
entities in $\cE$ and each column $\col$ to one relation in \cR. To this end,
we perform a sequence of five operations, described below.

\subsection{Step 1: Candidate Entity Selection}
\label{sec:candgen}
First, we identify the key-column (if any)
using the heuristics proposed by~\cite{Ritze2015MatchingHT}, which consists of
selecting the column with most unique non-numeric values breaking ties by
choosing the leftmost one. This heuristics works well in practice so we apply it
without modifications. Only the tables with valid key columns are considered
since these are the only ones for which we can (potentially) extract factual
knowledge.

For every cell in the key column, we then select a set of entity candidates. We represent this
computation with the function $\textsf{Cand}(\rho)$ which takes in input a generic row
$\rho$ and returns all entities in $\cE$ which are potential candidates with
$\rho$. This function is implemented by 1) indexing all the labels in $\cK$, 2)
retrieving the labels which contain the cell value of the key column, 3)
returning the entities associated to the labels.
Let $e \in \textsf{Cand}(\rho)$ be a potential entity candidate for row $\rho$. We call
the tuple $(\rho,e)$ a \emph{row-entity assignment}. If $\textsf{Cand}(\rho)$ is empty,
then $\rho$ is ignored. Otherwise, the table interpretation
process will determine which row-entity assignment should be selected.

\extended{
\begin{example}
   In the table (a) of figure \ref{fig:method}, we assume that the first column
   is the key column, because it is the leftmost column with non-numeric
   unique values. The label index returns a set of scored candidate entities per
   row (b).
\end{example}
}

The label matches are ranked using length-normalised smoothed TF-IDF.
In our case, the query corresponds to the cell
value of the key column, while the documents are all labels in $\cK$.
Identically to \cite{Ritze2015MatchingHT},
we (1) take only the first result if it is much better than the next and (2)
take the top three labels otherwise. The final set of candidates consists of all
entities associated with these labels.

Typically entities are explicitly linked to
labels with direct relations (e.g., \texttt{rdfs:label} ~\cite{rdfs}).
However, more links can be retrieved
if we also consider titles and disambiguation pages. In our approach, we add
also these labels to the index because we observed that this leads to a
substantial increase of the recall. At this stage, it is important to have a
high recall because the subsequent operations cannot recover in case we fail to
retrieve the correct mapping. In the definitions below, we denote these sets of
labels for each entity as $\textsf{Labels}(e)$.

\subsection{Step 2: Computation of the Priors} In this step, we compute a score
of the row-entity assignments by comparing all cell values in the row with all
the labels of entities that are connected to the candidate entities. To this
end, we first define attribute links, and related labels of an entity $e$ as
\begin{align} \textsf{Links}(e) =&~ \{\langle \rel,v \rangle ~|~ \langle
e,\rel,v \rangle \in \cF\} \\ \textsf{LinkLabels}(e,\rel) =&~ \{l ~|~ \langle
\rel,v \rangle \in \textsf{Links}(e), l \in \textsf{Labels}(v) \} \end{align}
Intuitively, $\textsf{Links}(e)$ contains all links of $e$ while
$\textsf{LinkLabels}(e,\rel)$ represents the labels at the other end of the
$\rel$-links from $e$. Then, we introduce the function
\begin{align} \textsf{Match}(c,\rho,e,r) =& \max_{s \in \textsf{Cell}(c,\rho)}~
\max_{l \in \textsf{LinkLabels}(e,\rel)} \textsf{TokenJaccard}(s,l) \end{align}
to compute the highest attainable string similarity between the cell at column
$c$ and row $\rho$ and the values of the $\rel$-links from $e$. Here,
$\textsf{Cell}(i,j)$ returns the content of the cell at row $i$ and column $j$
in a table with $n$ rows and $m$ columns, while $\textsf{TokenJaccard}$ is the
Jaccard index $J(A,B) = \frac{|A \cap B|}{|A \cup B|}$ of the tokens in each
string. For instance, in the table in Fig.~\ref{fig:method} each cell is matched
to each attribute of the corresponding row-entity candidates, e.g.,
$\textsf{Match}(3,4,\mathtt{The\_Producers\_(1968\_film)},\mathtt{director})$ is
the score that quantifies to what extent the content of the cell at coordinates
$(3,4)$ matches the string ``Mel Brooks'', which is the label of the director of
the film. Note that we treat the content of every cell as a string. There are some
approaches that use type-specific cell and column matching methods
\cite{Ritze2015MatchingHT,Pham,zhang2017effective,Limaye2010AnnotatingAS}, but a
combination of our method with these techniques should be seen as future work.


We can now compute likelihood scores for mapping cells to relations
(Eq.~\ref{eq:cellscore}), and for mapping columns to relations
(Eq.~\ref{eq:colscore}) to aggregate and normalise these scores on the row and
column levels respectively:
\begin{align}\label{eq:cellscore} \textsf{CellScore}(c,\rho,\rel) =&~ \frac{1}{
|\textsf{Cand}(\rho)| } \sum_{e \in \textsf{Cand}(\rho)}
    \textsf{Match}(c,\rho,e,\rel) \\ \label{eq:colscore} \textsf{ColScore}(c,\rel) =&~ \frac{ \sum_{i=0}^{n}
\textsf{CellScore}(c,\rho_i,\rel) } { \sum_{i=0}^{n} \sum_{\rel' \in \cR }
\textsf{CellScore}(c,\rho_i,\rel') } \end{align}

For instance, in Fig.~\ref{fig:method}a,
$\textsf{CellScore}(4,3,\mathtt{director})$ returns the likelihood that the cell
(4,3) matches the relation \texttt{director}, while
$\textsf{ColScore}(3,\mathtt{director})$ returns the aggregated scores for
column $3$ considering all rows in the table.

Since $\textsf{ColScore}(c,\rel)$ is the likelihood score that column $c$ maps to
relation $\rel$, we can use this value to construct the prior distribution of all
assignments to $c$. Furthermore, we can use these scores to refine the
likelihood of the possible row-entity matchings. We compute such likelihood as
\begin{align}
    \label{eq:rowscore}
    \textsf{RowScore}(\rho,e) =&~
        \frac{1}{m}  \sum_{i=0}^{m} \max_{\rel \in \cR} \textsf{ColScore}(c_i,\rel)
        \times \textsf{Match}(c_i,\rho,e,\rel)
\end{align}

In essence, Eq.~\ref{eq:rowscore} computes the likelihood of an entity-row
matching as the average best product that each cell matches to a certain
attribute $(\rel,e)$ ($\textsf{Match}(\cdot)$) with the general likelihood that the
column matches to $\rel$ ($\textsf{ColScore}(\cdot)$). We use the values of
$\textsf{RowScore}$ to build a prior distribution for all entity-row matches.

\subsection{Step 3: Entity Similarity Scores} Both prior distributions computed
with Eqs.~\ref{eq:cellscore} and ~\ref{eq:colscore} rely on the Jaccard Index.
Thus, they are distributions which are ultimately built on the string
similarities between the strings in the cells and the entities' labels. We use
these scores to compute similarity scores between pairs of candidate entities
across the rows. In the next step, we will use these similarities to compute
better entity-row likelihood scores than the ones of Eq.~\ref{eq:rowscore}.

First, we weigh all links $\langle \rel,v\rangle$ depending on their
popularities across the entities in the table and the corresponding prior of the
assignments that use them. To this end, we define the function
$\textsf{LinkTotal}$ as
\begin{align}
    \textsf{LinkTotal}(\rel,v) =&~
    \sum_{i=0}^{n} \max_{e \in \textsf{Cand}(\rho_i)}
    \textsf{RowScore}(\rho_i,e) [ \langle \rel,v
    \rangle \in \textsf{Links}(e) ]
\end{align}
where $[x]$ returns $1$ if $x$ is true or 0 otherwise. Note that
since $\textsf{RowScore}$ returns a value between 0 and 1,
$\textsf{LinkTotal}(\cdot)$ returns $n$ in the best case.

Then, we represent the coverage and saliency of $\langle r,v\rangle$ by
normalising the value $\textsf{LinkTotal}(r,v)$ with respect to the table and the KG:
\begin{align}\label{eq:cover} \textsf{Cover}(\rel,v) =&~ \frac{
    \textsf{LinkTotal}(\rel,v) }{\sum_{i=1}^{n}[\langle \rel,v\rangle \in \cup_{e\in
\textsf{Cand}(\rho_i)} \textsf{Links}(e)]} \\\label{eq:salience}
    \textsf{Salience}(\rel,v) =&~ \frac{ \textsf{LinkTotal}(\rel,v) }{ |\{e \in \cE ~|~
\langle \rel,v \rangle \in \textsf{Links}(e)\}|}
\end{align}
Intuitively, $\textsf{Cover}(\cdot)$ computes the popularity of $\langle \rel,v
\rangle$ among the rows of the table, while $\textsf{Salience}(\cdot)$ considers all
entities in $\cK$. We combine them as
\begin{equation}\label{eq:linkscore} \textsf{LinkScore}(\rel,v)
=~ \textsf{Cover}(\rel,v) \times \textsf{Salience}(\rel,v) \end{equation} so that we
can rank the attributes depending both on their coverage within the table and
popularity in the KG.
This combination allows us to give low ranks to
attributes, like $\langle$\texttt{isA,Resource}$\rangle$, which should not be
considered despite their high coverage since they are not informative. In
contrast, it can boost up the score of attributes with a medium coverage in case
they have a high saliency.

Finally, we use the scores from Eq.~\ref{eq:linkscore} to compute a
similarity score between pairs of entities. We compute
the similarity between entities $e_1$ and $e_2$ as
\begin{equation}\label{eq:e2e}
    \textsf{EntitySimilarity}(e_1,e_2) =
    \sum_{\langle \rel,v \rangle \in \textsf{Links}(e_1) \cap \textsf{Links}(e_2)}
    \textsf{LinkScore}(r,v)
\end{equation}

\subsection{Step 4: Disambiguation}

Now, we compute which are the row-entity assignments which maximise the
coherence in the table, i.e., maximise the similarity between the entities.
These assignments are determined using Loopy Belief Propagation (LBP)
\cite{Pearl1989ProbabilisticRI}.

We model each row-entity prediction as a categorical random variable, for which
the label score $\textsf{RowScore}(\rho,e)$ is the prior
distribution (Fig.~\ref{fig:method}d1). For convenience, we can view these
scores as a sparse matrix $\bm{L}$ of size $n \times |\mathcal{E}|$. The
variables are connected to each other with the edge potentials being defined by
entity-entity similarities $\textsf{EntitySimilarity}(e_1,e_2)$
(Fig.~\ref{fig:method}d2; equivalently represented by a matrix $\bm{S}$), which forms a complete graph. Since this graph has loops it is not possible to
perform exact inference. Therefore we approximate it by executing LBP.
Additionally, all our edge potentials are identical. This causes all nodes to
receive identical information from each other. Instead of having separate
messages for each node, we thus have a single vector-valued message
that provides the belief updates for our nodes:
\begin{align} q_e
    =&~ \prod^n_{\rho=0} \sum_{e' \in \textsf{Cand}(\rho)} L_{\rho,e'} \times
    S_{e,e'} =~ \prod^n_{\rho=0} (\bm{L}\bm{S})_{\rho, e} \\ \label{eq:lbp1}
    C_{\rho,e} =&~ L_{\rho,e} \times q_e
\end{align}
where $q_e$ indicates how similar entity $e$ is to all weighted candidates of
all rows, and $C_{\rho,e}$ is the coherence score of entity $e$ for row $\rho$
(Figs.~\ref{fig:method}e and~\ref{fig:method}f respectively). Because the main
operation consists of a single matrix multiplication, computation is fast and
can be parallelized by standard matrix processing libraries.

LBP can be run for multiple iterations (in our case, replacing $L_{\rho,e'}$ by
$C_{\rho,e'}$), but is not guaranteed to converge
\cite{Pearl1989ProbabilisticRI}. In
fact, we observed that sometimes an excessive number of iterations led to
suboptimal assignments. This occurred when the entity similarity scores
(Eq.~\ref{eq:e2e}) were not accurate due to missing attributes in the KG and
ended up ``overriding'' the more accurate priors that were computed considering
only label similarities (Eq.~\ref{eq:rowscore}) when they are combined in the
following step. From our experimental analysis, we observed that in the
overwhelming majority of the cases a single iteration of LBP was enough to
converge. Therefore, we apply Eq.~\ref{eq:lbp1} only once without further
iterations.

As we can see from Eq.~\ref{eq:lbp1}, the selection of the entity for row
$\rho$ relies on two components, $\bm{L}$ and $\bm{q}$: The first takes into
account to what extent the entity label matches the label of candidate entities
and to what extent the labels of the attributes matches with the remaining cell
values. The second considers the coherence, i.e., the mappings that
maximise the similarity between the entities.

Finally, we disambiguate rows by choosing the highest-rated candidate
$\hat{e}_\rho = \mathrm{argmax}_{e}~C_{\rho,e}$. Then, we re-calculate
$\textsf{ColScore}(c,r)$ with the updated set of candidates containing only the
predicted entity $\textsf{Cand}(\rho)=\{\hat{e}_\rho \}$ and disambiguate
columns by choosing the highest scoring relation $ \hat{r}_c =
\mathrm{argmax}_{\rel}~ \textsf{ColScore}(c,r) $. After this last step is
computed, our procedure has selected one entity per row and one relation per
attribute column. In the next section, we discuss how we can extract triples
from the table.




 \label{method}

\section{Slot-Filling}
After the table is interpreted, we can extract partial triples of the form
$\langle \sub,\rel,?\rangle$ where $\sub$ are the entities mapped to rows and $\rel$ are the
relations associated to columns. If the cell contains numbers or other
datatypes (e.g., dates) that we can add the cell value to the KG as-is, but this
is inappropriate if the content of the cell refers
to an entity. In this case, we need to map the content of the cell to an entity
in the KG.


The disambiguation of the content of a cell could be done by querying our label index
precisely the same way as done in Sec.~\ref{sec:candgen}. However, this
extraction is suboptimal since now we have available some context, i.e.,
$\langle \sub,\rel,?\rangle$ that we can leverage to refine our search space. To
this
end, we can exploit techniques for predicting the likelihood of triples given
the KG's structure, namely KG embeddings provided by the TransE
algorithm~\cite{transe}.
Given in input $e_i$, i.e., the entity associated to row $i$ and
    $\rel_j$, i.e., the relation associated to column $j$, our goal is to
    extract
    a fact of the form $\langle e_i,\rel_j,x\rangle$ where entity $x$ is
    unknown.
We proceed as follows:

\squishlist
\item[1.] We query the label index with the content of $\textsf{Cell}(i,j)$ as
    done for the computation of $Cand(\cdot)$ in Sec.~\ref{sec:candgen}. This
    computation returns a list of entity candidates $\langle
    e_1,\ldots,e_n\rangle$ ranked based on label string similarities.

\item[2.] For each candidate $e_k \in \langle e_1,\ldots,e_n\rangle$, we compute
    $\textsf{Rank}(k)=d(\bm{e_i} + \bm{r_j}, \bm{e_k})$ where $d$ is the
    distance measure used to compute the TransE embeddings (we use the $L_1$
    norm), and $\bm{e_i,r_j,e_k}$ are the TransE vectors of $e_k,\rel_j,e_i$
    respectively.

\item[3.] We return $\langle e_i, r_j, e_k\rangle$ where $e_k$ is the entity
    with the lowest $\textsf{Rank}(k)$, i.e, has the closest distance hence it is the
    triple with the highest likelihood score.
\squishend

 \label{slotfilling}
\section{Evaluation} 

We implemented our method into a system called \texttt{TAKCO} (TAble-driven KG
COmpleter). The code is available
online\footnote{\url{https://github.com/karmaresearch/takco}}.

\extended{
Our implementation uses two
additional systems:
\emph{Trident}\footnote{\url{https://github.com/karmaresearch/trident}}, an
in-house triple store to query the KG; and \emph{Elasticsearch 6.4.2}, a
well-known system used for building and querying the label index. Moreover, we
reimplemented the TransE algorithm for creating the KG embeddings. Since our KGs
contain millions of nodes and edges, we parallelized the learning using
Hogwild!~\cite{hogwild} (we empirically verified that this form of parallelism
does not affect the quality of the embeddings).
}


\leanparagraph{Baselines} Since our goal is to extract novel facts from tables,
we considered existing systems that perform slot-filling as baselines. In
particular, we considered the systems \tkmatch~\cite{Ritze2015MatchingHT} and
\tableminer~\cite{zhang2017effective} because of their state-of-the-art results.
There are other systems that implement only parts of the pipeline, for instance
entity disambiguation (see Section~\ref{sec:related} for an overview). An
important system in this category is TabEL~\cite{Bhagavatula2015TabELEL}, which
exploits co-occurrences of anchor links to entity candidates on Wikipedia pages
for predicting a coherent set of entities. Although such system can potentially
return better performance on entity disambiguation, we did not include it in our
analysis due its reliance on additional inputs. A comparison between the
performance of our method for the subtask of entity disambiguation, and more
specialized frameworks like TabEL should be seen as future work.

The system \tkmatch implements a series of matching steps that match table rows
to entities, using similarities between entity property values and the table
columns. The \tableminer system consists of two phases that are alternated until
a certain confidence level has been reached. Note that these approaches
primarily focus on table interpretation. In contrast, we provide an end-to-end
system which considers also the operation of slot-filling.

The first system is designed to work with a specific subselection of
DBpedia~\cite{auer2007dbpedia} while the second system was originally built to
use the Freebase API. We have performed some slight modifications to their
source code so that we could perform a fair comparison. For \tkmatch, we
modified the system to be able to use an augmented set of candidates so that in
some experiments we could measure precisely the performance of table
interpretation. For \tableminer, we modified the system so that we could use
different KGs without API access.

\leanparagraph{Knowledge Graphs} Our method can work with any arbitrary KG. We
consider DBpedia (so that we could compare against \tkmatch) which is a popular
KGs created from Wikipedia and other sources. We use two versions of DBpedia:
The first is the triple-based version of the tabular subset used by \tkmatch.
This is a subset of DBpedia from 2014 and we consider it so that we can
perform an exact comparison. It contains 3.4M entities and 28M facts.
Additionally, we also use the latest version of the full KG (version 2016-10). The
full DBpedia contains 15M entities (including entities without labels and
redirected entities) and 110M facts.
Finally, we compare our performance using Wikidata (``truthy'' RDF export,
acquired on Oct 2018), which
has 106M entities and 1B facts. For evaluation, we map the gold standard to
Wikidata using $\texttt{owl:sameAs}$ links from DBpedia.

\leanparagraph{Testsets} To the best of our knowledge, there are two
openly available datasets of tables that have been annotated for the purpose of
table interpretation. The first one is the \emph{T2D}
dataset~\cite{Ritze2015MatchingHT}, which contains a subset of the WDC Web
Tables Corpus -- a set of tables extracted from the CommonCrawl web
scrape\footnote{\url{http://webdatacommons.org/webtables/}}. We use the latest
available version of this dataset (v2, released 2017/02). In our experiments, we
disregarded tables without any annotation. The resulting dataset contains
238 entity tables with 659 column annotations and 26106 row-entity
annotations. Throughout, we refer to this dataset as \emph{T2D-v2}.

The second dataset is \emph{Webaroo}, proposed by~\cite{Limaye2010AnnotatingAS}.
Tables in this dataset were annotated with entities and relations in YAGO.
While these tables are a less varied sample of the ones in the T2D, they allow
us to study the behaviour of the systems on a dataset with different
annotations. This dataset contains 429 entity tables with 389 and 4447 column
and row-entity annotations respectively. In order to test the performance of
\tkmatch with this dataset, we ``ported'' the YAGO annotations to DBpedia using
the Wikipedia links they refer to. Finally, we tested the scalability of our
system by running it on a large set of Wikipedia tables
\cite{Bhagavatula2015TabELEL}. Instructions to obtain these datasets are
available in the code repository of our system.



\begin{figure*}[ht]
    \centering
    \small
    \subfloat[Performance tradeoff, T2D-v2\label{fig:instance_t2dv2}]
    {\includegraphics[width=0.49\textwidth, trim={0 0 0 0},clip]
    {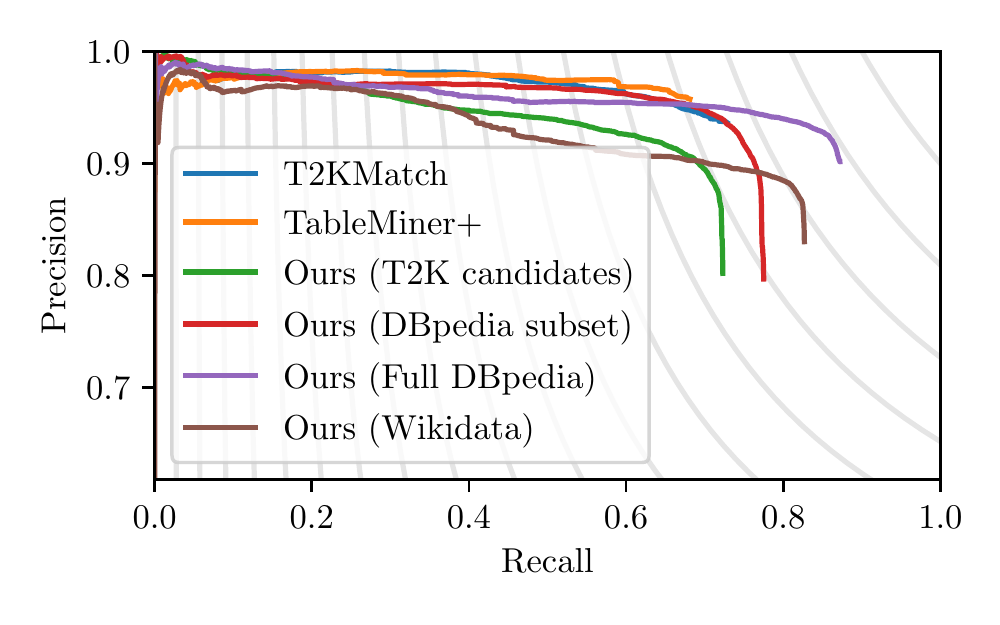}}
    \subfloat[Performance tradeoff, Webaroo\label{fig:instance_webaroo}]
    {\includegraphics[width=0.49\textwidth, trim={0 0 0 0},clip]
    {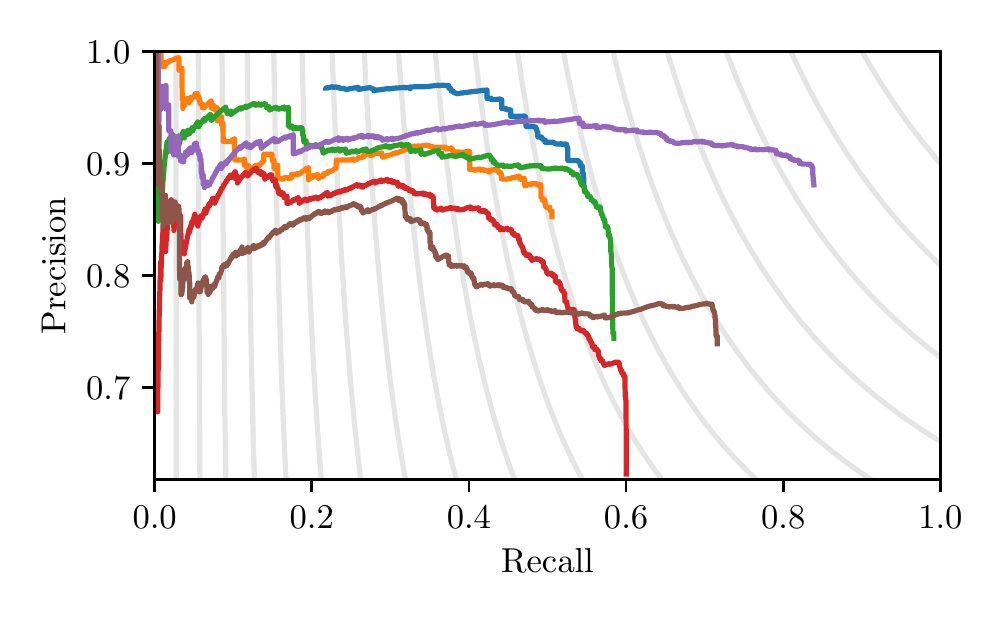}}\\
    \subfloat[Row-entity evaluation, T2D-v2 \label{tab:instance_t2dv2}]{
    \setlength{\tabcolsep}{0.3em} \begin{tabular}{lrrrr}
\toprule
                System &       Pr. &       Re. &     $F_1$ \\
\midrule
              T2KMatch &       .94 &       .73 &       .82 \\
           TableMiner+ &  \bf{.96} &       .68 &       .80 \\
 Ours (T2K candidates) &       .88 &       .72 &       .79 \\
 Ours (DBpedia subset) &       .90 &       .76 &       .83 \\
   Ours (Full DBpedia) &       .92 &  \bf{.86} &  \bf{.89} \\
       Ours (Wikidata) &       .87 &       .82 &       .84 \\
\bottomrule
\end{tabular}
}
    \subfloat[Row-entity evaluation, Webaroo\label{tab:instance_webaroo}]{
    \setlength{\tabcolsep}{0.3em} \begin{tabular}{lrrrr}
\toprule
                System &       Pr. &       Re. &     $F_1$ \\
\midrule
              T2KMatch &       .88 &       .55 &       .67 \\
           TableMiner+ &       .85 &       .51 &       .63 \\
 Ours (T2K candidates) &       .74 &       .58 &       .65 \\
 Ours (DBpedia subset) &       .72 &       .59 &       .65 \\
   Ours (Full DBpedia) &  \bf{.88} &  \bf{.84} &  \bf{.86} \\
       Ours (Wikidata) &       .77 &       .71 &       .74 \\
\bottomrule
\end{tabular}
}

    \caption{Row-entity evaluation scores and precision-recall tradeoff for the
    T2D-v2 and Webaroo datasets (the isolines of constant $F_1$ score are shown
in grey). Precision, recall, and $F_1$ are calculated at the threshold of
maximum $F_1$.}
    \label{fig:eval_tableinter}
\end{figure*}

\subsection{Table Interpretation}

\begin{figure*}[ht]
    \centering
    \small
    \raisebox{-.5\height}{\includegraphics[width=0.49\textwidth, trim={10 12
    10 11},clip]{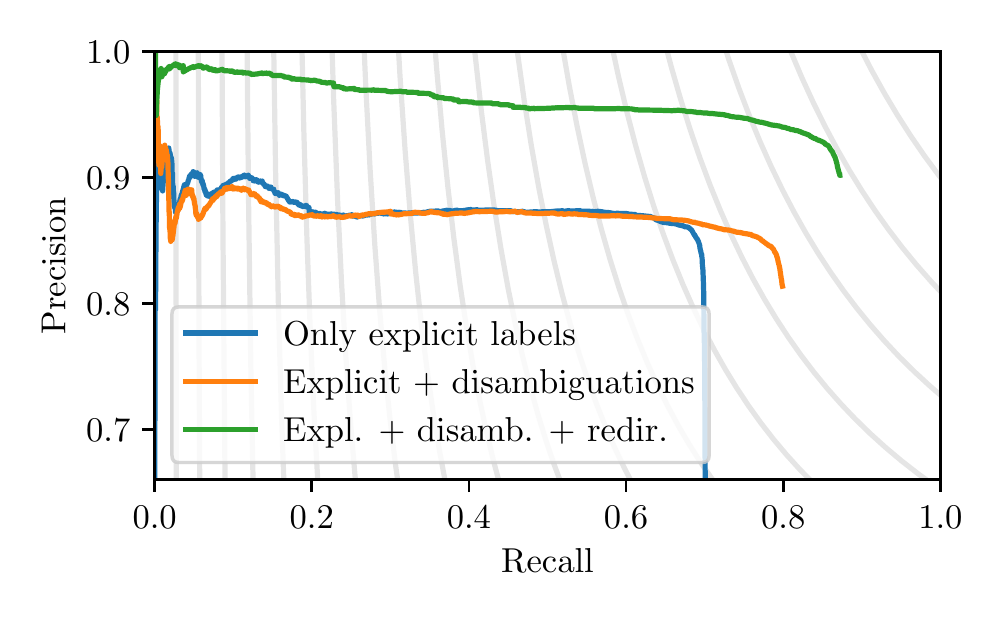}}
    \setlength{\tabcolsep}{0.25em} \begin{tabular}{lrrrr}
\toprule
                     System &       Pr. &       Re. &     $F_1$ \\
\midrule
       Only explicit labels &       .85 &       .69 &       .76 \\
 Explicit + disambiguations &       .84 &       .79 &       .81 \\
   Expl. + disamb. + redir. &  \bf{.92} &  \bf{.86} &  \bf{.89} \\
\bottomrule
\end{tabular}

    \caption{Row-entity evaluation scores and precision-recall tradeoff of
    our approach given different label sources, on T2D-v2.}
    \label{fig:eval_tableinter2}
\end{figure*}

We evaluate the performance of determining the correct row-entity assignments,
which are the key output for table interpretation.
Fig.~\ref{fig:eval_tableinter}b,d and Fig.~\ref{fig:eval_tableinter}a,c report
a
comparison of the performance of our method against the baselines. We
measure the precision/recall tradeoff (obtained by altering the threshold value
for accepting mappings), and precision, recall, and $F_1$ (shown at the
threshold of maximum $F_1$) on all predictions. The precision decreases whenever
a system makes a wrong
prediction while the recall is affected when no entity is selected. Not
predicting a match for a row can have several causes: the candidate set for that
row might have been empty, the annotated entity might not have been in the KG
(this occurs when we use a subset), or when all candidates have been pruned away
during the interpretation (this occurs with the baselines).

In these experiments, we configured our system with three different settings:
First, we use the same KG and the candidates (i.e., the output of
$\textsf{Cand}(\cdot)$) used by the other two systems. We refer to this setting
as ``T2K candidates''. Then, we use the KG subset used by \tkmatch in our own
label index and disambiguation (``DBpedia subset''). Finally,
we use our own candidate set generation and full KG (``Full DBpedia'').
By evaluating the performance of our method with these settings, we can compare
the performance of our approach given the limitations of the inputs that the
other systems face.

From the results reported in the figures, we can make a few considerations.
First, our method returns a comparable recall but an inferior precision
than the baselines if we use the set of candidates from \tkmatch, but is able to
match its performance in terms of $F_1$ when using the same KG.
However, the baselines are limited with respect to KGs. In fact, \tkmatch
requires that DBpedia is translated into
a tabular format while our method does not have this restriction. If our method
is configured to use the full DBpedia KG, then it returns the highest recall
with only a
small loss in terms of precision. This translates in a significantly higher
$F_1$ score than the best of the baselines. These are
positive results since a high recall is important for extracting novel facts.

While the precision of our system is low in the limited-input setting, many of
the errors that it makes are due to problems with the candidate set and the KG.
Therefore, we evaluated a scenario (not shown in the figures of this paper) in
which we artificially expanded the candidate set to always include the gold
standard. This means we are artificially making use of a ``perfect'' candidate
index. Even with this addition, \tkmatch is unable to use these
candidates for prediction and returns the same results. In contrast, manually
adding them to our system leads to both a notably higher recall and precision.

This indicates that our method is sensitive to the candidate generation, i.e.,
to the very first selection of candidates using the index label. To evaluate how
our system behaves with richer label indices, we evaluated our method on T2D-v2
with three different label indices. The first index
only uses the explicit labels of the entities. The second one includes also the
labels that we obtain from redirect pages in Wikipedia. The third one adds also
the labels we obtain from the disambiguation pages. The results of this
experiment are reported in Fig.~\ref{fig:eval_tableinter2}. As we can see
from these results, including more labels per entity significantly improves both
the precision and recall of our system.



\begin{figure}[h]
\centering
\subfloat[The scores for extracting novel and redundant triples from T2D-v2,
measured at the acceptance threshold of maximum $F_1$. \label{tab:triple_t2dv1}]
{
\setlength{\tabcolsep}{0.5em}
\begin{tabular}{l|rrr|rrr}
                System & \multicolumn{3}{c|}{Redundant} & \multicolumn{3}{c}{Novel} \\
                       &       Pr. &       Re. &     $F_1$ &       Pr. &       Re. &     $F_1$ \\
\hline
              T2KMatch &       .84 &       .82 &       .83 &  \bf{.76} &       .66 &       .71  \\
          TableMiner+ &  \bf{.86} &       .73 &       .79 &       .73 &       .56 &       .63  \\
 Ours (T2K candidates) &       .81 &       .84 &       .83 &       .61 &       .71 &       .66  \\
 Ours (DBpedia subset) &       .83 &       .90 &       .86 &       .59 &       .76 &       .66  \\
   Ours (Full DBpedia) &       .83 &  \bf{.96} &  \bf{.89} &       .70 &  \bf{.83} &  \bf{.76}  \\
\end{tabular}

}\\
\subfloat[The precision-recall tradeoff
curve on T2D-v2.]{\includegraphics[width=0.99\textwidth, trim={0 0
    0 0},clip]{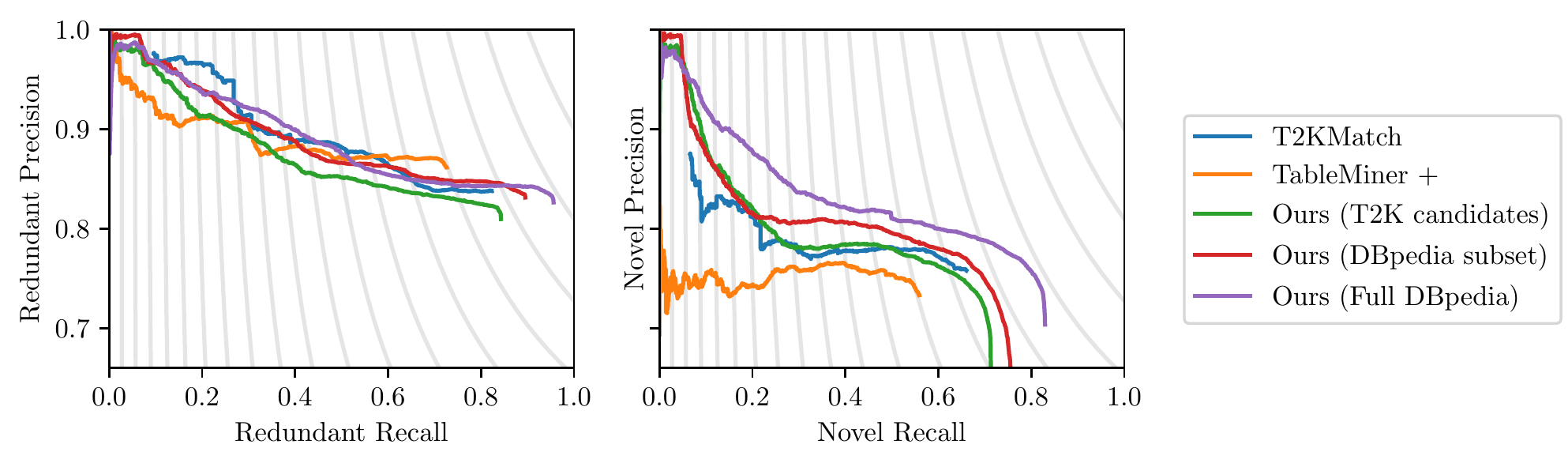}}
\caption{The novel and redundant precision-recall tradeoff for the T2D-v2
dataset (in gray, the isolines of constant $F_1$ score). Unlike the experiments
in the previous figures, here the bias towards extracting known (redundant)
facts is made explicit and we focus on finding novel KG facts in web tables.}
\label{fig:triple_t2dv2}
\end{figure}

\subsection{Measuring Redundancy}

Current systems (e.g.,~\cite{Ritze2015MatchingHT,zhang2017effective}) were
evaluated against a set of manual annotations, and scored on the individual
subtasks of table interpretation. Such evaluation did not consider the
novelty of facts that the system has extracted. In other words, no difference
was made between predictions of already known facts or new knowledge, but this
difference is important in our context.  In order to fill this gap, we need to
distinguish between these cases when measuring performance.

Given in input a KG $\cK=(\cE,\cR,\cF)$, an extraction technique like ours is
expected to yield a new set of predicted facts $\cF_P$ over $\cE$ and $\cR$ from
an input source like web tables. If we have gold standard table annotations, we
can generate another set of facts $\cF_G$ and use them for evaluating how many
facts in $\cF_P$ are correct. Note that both $\cF_P$ and $\cF_G$ might contain
facts that are either in $\cF$ or not. So far, current techniques have been
evaluated w.r.t. the set of true positives $\mathcal{F}_G \cap \mathcal{F}_P$
(correctly extracted facts) and false negatives as $\mathcal{F}_G \setminus
\mathcal{F}_P$ (valid facts that were missed). These measures do not take the
redundancy of the extracted facts into account, while the redundant information
exceeds the novel information for benchmark datasets \cite{kruit2018extracting}.

In Fig.~\ref{fig:triple_t2dv2}, we show the evaluation of the correctness of
\emph{novel} and \emph{redundant} facts separately. Crucially, our system
significantly outperforms the baselines with respect to the recall of
novel facts, which is paramount to KG completion.
In the tradeoff curve for novel triples (Fig.~\ref{fig:triple_t2dv2}b), we also
outperform the state-of-the-art regarding precision for most threshold values.

\subsection{Slot-filling}

To test the scalability of our system, we have run it on all 1.6M tables in the
Wikitable dataset. The first step concerns detecting entity tables with key
columns that contain entity labels. This process returned 786K tables. Then, we
proceeded with the retrieval of entity candidates. About 288K tables did not
contain any entity in DBpedia, thus were discarded. The table interpretation
process was launched on the remaining 498K tables. Our approach is trivially
parallelizable, and runs in 391 ms per table on average.

\begin{table}
\centering
\setlength{\tabcolsep}{0.5em}
\begin{tabular}{llcc}
\toprule
Ranking & Dataset & Prec@1 & Prec@3 \\
\midrule
Only Label Index (TF-IDF score) & Wikitable & 0.37 & 0.42 \\
& T2D-v2 & 0.24 & 0.31 \\
Labels + Embeddings (TransE) & Wikitable & $\bm{0.61}$ & $\bm{0.72}$ \\
& T2D-v2 & $\bm{0.62}$ & $\bm{0.74}$ \\
\bottomrule
\end{tabular}

\vspace{1mm}
\caption{Precision of slot-filling with/out KG embeddings, calculated on
redundant extractions.}
\label{tab:completion}
\end{table}





From these tables, we extracted 2.818.205 unique facts for 1.880.808 unique
slots of the form $\langle \sub,\rel,?\rangle$. Of those slots, 823.806 already
contain at least one entity $\obj$ in the KG. However, we do not know whether our
extractions are redundant, or $t$ represents a new extraction that should be
added to the existing ones in the KG. To determine the novelty, we queried the
label index for every extracted fact and discovered that in 307.729 cases the
labels were matching. We can assume these extracted facts to be redundant. From
these numbers, we conclude that our extraction process has produced about 1.6M
extractions for which we have no evidence of redundancy and thus can be
considered as novel. A manual analysis over a sample confirmed this conclusion.

Finally, we evaluated the effectiveness of our procedure for re-ranking the
extractions using the KG embeddings on the Wikitable and T2D-v2 datasets. To
this end, we compare the na\"ive index-based ranking obtained by simply picking
the top result returned from the label index against our strategy or re-ranking
considering the distance of the corresponding embeddings
(Sec.~\ref{slotfilling}). We chose to measure the precision for the first or
top-3 ranked candidates since this is a standard metric used to evaluate the
performance of link prediction~\cite{nickel2016review}.

Since we need to know the correct entity, we restricted this analysis to the
redundant extractions (i.e., the ones already in the KG) and disregarded the
novel ones.  Tab.~\ref{tab:completion} reports the results both when we consider
only the best result and the top three. We see that that our embedding-based
ranking outperforms the index-based ranking in both cases, and predicts the
correct entity at the top of the ranking in 61\% of the time, compared to 37\%
for the Wikitable dataset. Moreover, the relatively low results obtained with
the index-based ranking strategy indicate that labels are in general not
reliable for disambiguating attributes.
 \label{evaluation}

\section{Related Work} 
    \label{sec:related}

    The first system for interpreting web tables was introduced by Limaye et
    al.~\cite{Limaye2010AnnotatingAS}. The system uses a probabilistic graphical
    model that makes supervised predictions based on a large number of features.
    Subsequent work approached the problem with a task-specific knowledge graph
    \cite{Venetis2011RecoveringSO,Wang2012UnderstandingTO} and sped up
    predictions by limiting the feature set~\cite{Mulwad2013} or using
    distributed processing~\cite{Hassanzadeh2015UnderstandingAL}. Others used an
    entity label prior from hyperlinks on the web~\cite{Bhagavatula2015TabELEL},
    and interpreted tables in limited domains~\cite{Ran2015}.

    A separate family of table interpretation systems limit themselves to
    attribute matching.  The simplest approaches perform string similarities
    between the column headers and relation names or cell values and entity
    labels \cite{Efthymiou2016}. When no overlap between the table and KG can be
    assumed at all, the work at~\cite{Pham} uses supervised models based on
    features of the column header and cell values. Some approaches focus on
    matching tables to relation from Open Information Extraction
    \cite{Venetis2011RecoveringSO,Wang2012UnderstandingTO} or exploit
    occurrences of cell value pairs in a corpus of text
    \cite{Sekhavat2014KnowledgeBA,Cannaviccio2018}, and others perform
    supervised learning \cite{Pham,Ermilov2016}. While several approaches have
    been proposed that are limited to entity linking
    \cite{Efthymiou2017,Bhagavatula2015TabELEL}, the focus of our work is to
    optimize table interpretation for novel fact extraction.

    The systems evaluated in this paper are designed for open-domain table
    interpretation. In closed-domain settings, assumptions can reduce the
    redundancy of extractions. For example, the work of \cite{Ran2015} models
    the incompleteness in the domain subset of the KG by estimating class
    probabilities based on relations between entities, which the limited domain
    makes tractable. The systems of \cite{Wang2012UnderstandingTO} and
    \cite{Venetis2011RecoveringSO} use a probabilistic KG created from a web
    corpus for supporting table search. This type of KG offers many strategies
    for improving the recall of new knowledge because it allows for an explicit
    model of low-confidence facts.

    Several models use large web text corpora in addition to the information
    from the KG. The work by Bhagavatula et al.~\cite{Bhagavatula2015TabELEL}
    uses the anchor text of hyperlinks on the web to create a prior for instance
    matching that takes the popularity of entities into account. Additionally,
    it exploits co-occurrences of anchor links to entity candidates on Wikipedia
    pages for predicting a coherent set of entities. The work of
    \cite{Sekhavat2014KnowledgeBA} creates a set of syntactic patterns from the
    ClueWeb09 text corpus featuring entities from relations in the KG.  Given
    query pairs of entities from tables, the syntactic patterns from text
    featuring the query pair are matched to the patterns in the set. A
    probabilistic model then allows for the prediction of relations from the KG.
    A similar approach is taken by \cite{Cannaviccio2018}, who use a language
    model instead of extracted syntactic patterns. This approach queries a
    search engine with the entity pair, and classify the text that occurs
    between the entity mentions. A separate direction is the matching of numeric
    columns, either with metrics for numeric distribution similarity
    \cite{Neumaier2016} or sophisticated ontologies of quantities and
    statistical models \cite{Ibrahim2016}.


The survey at~\cite{ji2011knowledge} discusses approaches and challenges to the
slot filling task in the context of textual information extraction. Most systems
use distant supervision for textual pattern learning, and some employ cross-slot
reasoning to ensure the coherence of multiple extracted values.  Recently, work
on Universal Schemas by Riedel et al.~\cite{Riedel2013RelationEW} has allowed
the joint factorisation of textual extractions and KB relations and this boosts
slot-filling precision.

    In the field of data fusion, systems explicitly aim for high recall and use a post-processing filter to improve precision.
    In \cite{Munoz2014}, the extracted facts are filtered using machine learning models, and in \cite{Dong2014} they are filtered using a sophisticated statistical model of the KG.
    In \cite{Ritze2016Profiling}, the system of \cite{Ritze2015MatchingHT} is used to interpret a large collection of web tables, after which the extracted facts are filtered using several strategies.
    However, only 2.85\% of web tables can be matched, which is attributed to a topical mismatch between the tables and the KG.
 \label{related}

\section{Conclusion} 

We investigate the problem of extending KGs using the data found in Web tables.
Existing approaches have focused on overall precision and recall of facts
extracted from web tables, but it is important for the purpose of KG
completion that the extraction process returns as many (correct) {\em novel}
facts as possible.

We developed and evaluated a new table interpretation method to counter this
problem. Our method uses a flexible similarity criterion for the
disambiguation of entity-row matches, and employs a PGM to compute new
likelihood scores depending on how the various candidates are similar to each
other to maximise the coherence of assignments.
Because it combines the syntactic match between the tables and the KG with the
coherence of the entity predictions, it can confidently predict more candidates
for which the attributes in the table are not yet in the KG. Consequently, it
extracts more novel facts for KG completion.
For the task of slot-filling, we introduced a novel approach for attribute
disambiguation based on KG embeddings, which outperforms a naive label-based
approach.

We compared our method to two state-of-the art systems, and performed an
extensive comparative evaluation on multiple knowledge bases. Our evaluation
shows that our system achieves a higher recall during the interpretation
process, which is necessary to extract novel information. Furthermore, it is
able to extract more (correct) facts that are not yet in the KG.

Interesting directions for future work include the development of extensions for
tables where the entity is identified by multiple columns or where rows do not
necessarily describe entities. In particular, the heuristics for determining the
key column of the table (and whether such a column is present) would need to be
replaced by a model that reliably detects the type of table. Moreover, the
inclusion of external sources can be useful to extract more novel information
from the table. Finally, despite the remarkable work by different research teams
to produce good benchmark datasets, there is still the need for larger and more
diverse benchmarks to further challenge the state-of-the-art.
 \label{conclusion}

\bibliographystyle{splncs04}
\bibliography{biblio}

\end{document}